\documentclass[twocolumn]{aastex631}

\newcommand\asec{^{\prime\prime}}
\usepackage{graphicx}
\usepackage{color}

\begin{document}

\title{An Integral Field Unit for the Binospec Spectrograph}

\author[0000-0002-1311-4942]{Daniel Fabricant}
\email{dfabricant@cfa.harvard.edu}\affiliation{Center for Astrophysics | Harvard and Smithsonian\\
60 Garden Street\\
Cambridge, MA 02138, USA}
\author[0000-0001-6760-3074]{Sagi Ben-Ami}
\affiliation{Weizmann Institute of Science\\
Faculty of Physics\\
Rehovot, Israel\\}
\author[0000-0002-7924-3253]{Igor V. Chilingarian} \author{Robert Fata} \author[0000-0002-9194-5071]{Sean Moran} 
\author[0000-0001-8120-7457]{ Martin Paegert} \author{Matthew Smith} \author{Joseph Zajac}
\affiliation{Center for Astrophysics | Harvard and Smithsonian\\
60 Garden Street\\
Cambridge, MA 02138, USA}

\begin{abstract}
Binospec is a wide-field optical (360 to 1000 nm) spectrograph commissioned at the MMT 6.5m telescope in 2017.  In direct imaging mode Binospec addresses twin 8$^\prime$ (wide) by 15$^\prime$ (slit length) fields of view. We describe an optical fiber based integral field unit (IFU) that remaps a 12$^{\prime\prime}$ x 16$^{\prime\prime}$ contiguous region onto two pseudo slits, one in each Binospec channel.  The IFU, commissioned in 2023, fits into the space of a standard slit mask frame and can be deployed as desired in a mixed program of slit masks, long slits, and IFU observations.  The IFU fibers are illuminated by a hexagonal lenslet array with a 0.6$^{\prime\prime}$ pitch. A separate bundle of sky fibers consists of close-packed bare fibers arranged within an 11.8$^{\prime\prime}$ circular aperture. The 640 IFU fibers and 80 sky fibers have a core diameter of 150$\mu$m, corresponding to 0.90$^{\prime\prime}$. Three gratings are available, 270lpm with R$\sim$2000, 600lpm with R$\sim$5300, and 1000 lpm with R$\sim$6000.
\end{abstract}

\keywords{Spectroscopy --- Astronomical Optics}

\section{Introduction} \label{sec:intro}

Binospec \citep{2019PASP..131g5004F} has been in operation at the f/5 focus of the 6.5m MMT telescope since early 2018.  Binospec was designed for rapid survey spectroscopy and imaging at wavelengths between 360 and 1000nm with twin 8$^\prime$ (wide) by 15$^\prime$ (slit length) fields of view.  Binospec was designed to use exchangeable aperture masks but its long slit length (totaling 1800$^{\prime\prime}$ or $\sim$300mm also lends itself to use with an optical fiber integral field unit (IFU).  For operation with Binospec the main challenge is designing and assembling a compact IFU that fits within the confines of the aperture mask frames $\sim$260 by 290mm and $\sim$25mm thick to fit within a single aperture mask slot in the existing mask exchange mechanism.  The other principal design goal is to use a hexagonal lenslet array to provide continuous spatial coverage with minimal gaps.  \citet{2002PASP..114..892A} describe an early example of such an IFU for the Gemini GMOS spectrographs.  \citet{2000SPIE.4008..510P} describe a similar IFU for VIMOS on ESO's VLT and \citet{2004SPIE.5492..624S} an IFU for IMACS at Magellan.  There are many examples of IFUs with dedicated spectrographs, see e.g.~the introduction in \citet{2023arXiv231114230R}. Designing an IFU to coexist with multiple operating modes of an imaging spectrograph is a specialized problem due to space constraints.

\section{IFU Assembly and Design} \label{IFU}
\begin{figure}[b!]
\includegraphics[width=\hsize]{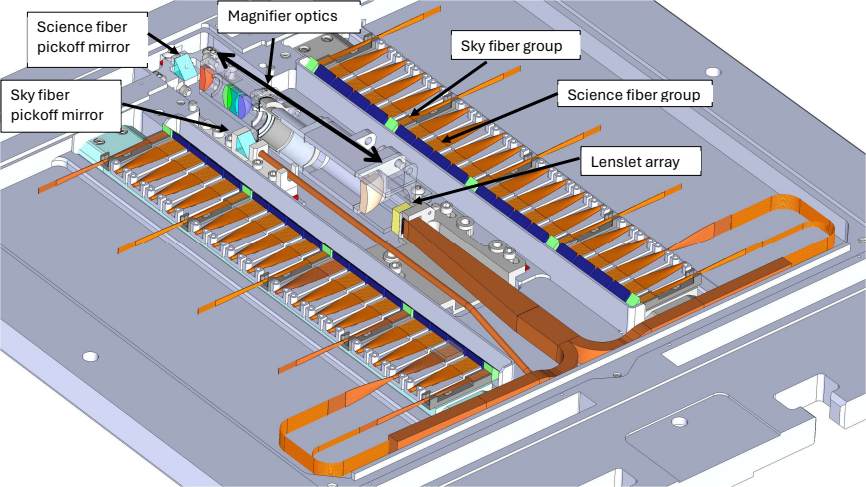}
\caption{Drawing showing the basic layout of the IFU frame. Science target light enters at the location labeled ``Science fiber pickoff mirror'', and passes through the magnifier optics and lenslet array. Off-target light for sky background subtraction enters at its own pickoff mirror, but passes through without additional focusing optics. Fibers are shown in orange, and are arranged in bundles of 20 (science) or 8 (sky), which end at 90-degree prisms arranged linearly on each side (shown blue for science and green for sky)}.
\label{fig:frame}
\end{figure}

The 12$^{\prime\prime}$ x 16$^{\prime\prime}$ IFU region may be completely filled by an extended source so we provide an array of 80 sky fibers offset by $\sim$50mm or $\sim$300$^{\prime\prime}$ from the center of the IFU field.  The fibers in the IFU and sky fiber array are oriented parallel to the aperture mask frame for compactness and view the sky through small fold mirrors. Prisms at the output end of the optical fibers direct light into the spectrograph as a pseudo slit at the correct angle and focal position to reproduce direct illumination from the MMT's f/5 optics and the concave focal surface \citep{2004SPIE.5492..767F}.  The sky and IFU fibers are type Polymicro FBP150170195 with 150$\mu$m cores, corresponding to 0.90$^{\prime\prime}$ at the input of Binospec.

\begin{figure}[t!]
\includegraphics[width=\hsize]{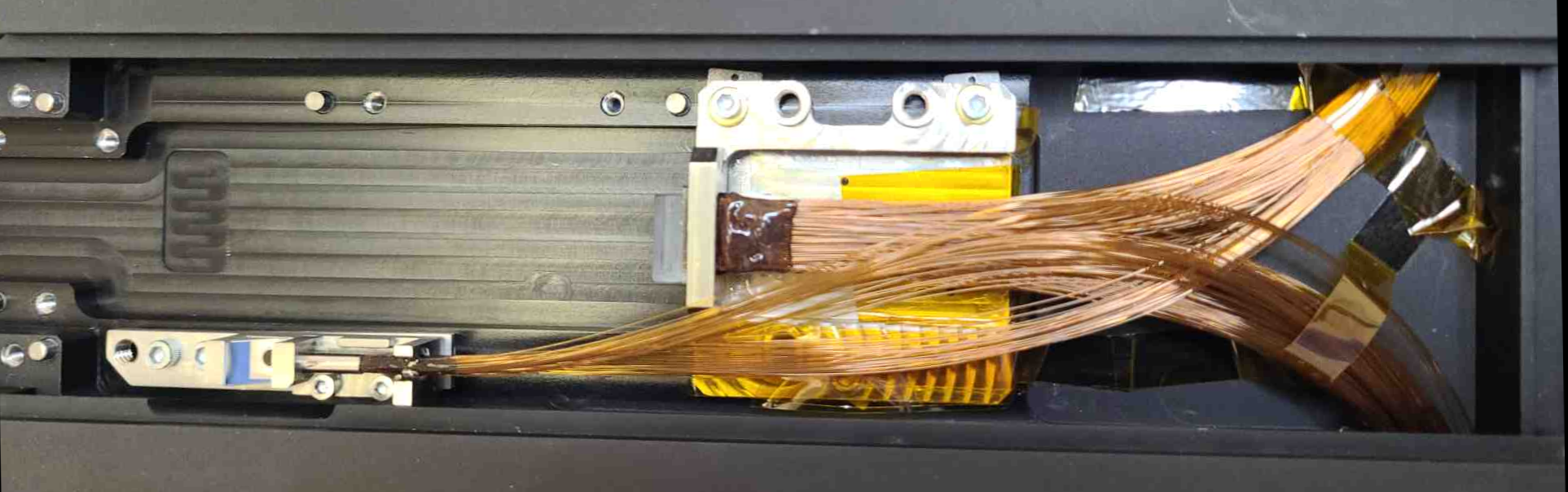}
\caption{Fibers leaving the sky bundle (lower left) and lenslet array (center).  The magnifier optics are removed in this photo.}
\label{fig:fibers}
\end{figure}

\begin{figure}[hb!]
\centering
\includegraphics[width=\linewidth]{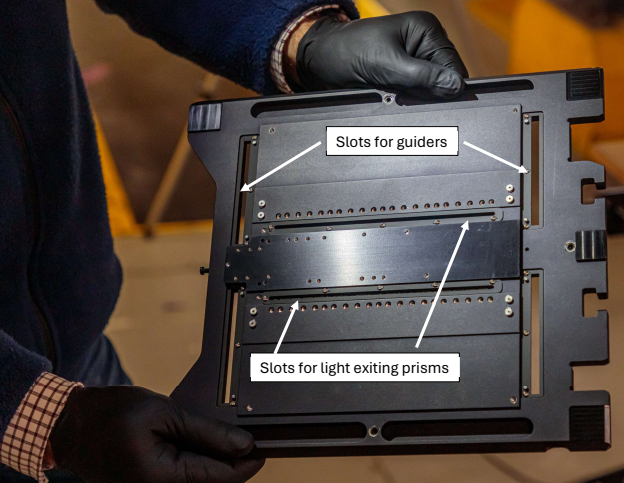}
\caption{Back surface of IFU assembly. All covers are in place. Light from the fiber bundles exits along the horizontal slits (via the 90 degree prisms), into the spectrograph optical path. The vertical open slots are windows through which the guide cameras can view guide stars. }
\label{fig:frame_back}
\end{figure}

\begin{figure}[ht!]
\centering
\includegraphics[width=\linewidth]{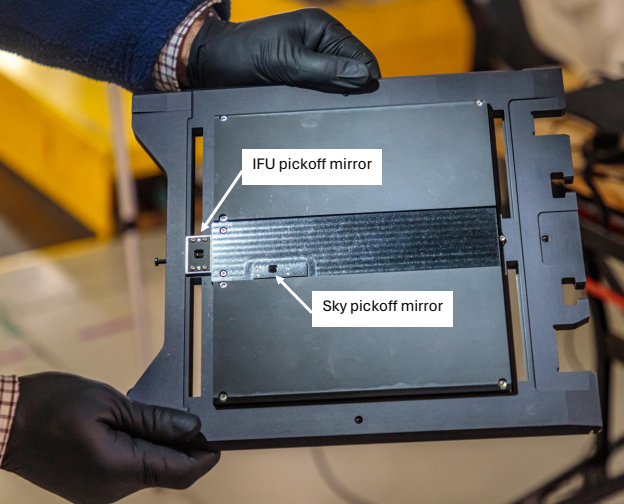}
\caption{Front surface of IFU assembly.  All covers are in place. The pickoff mirrors for the IFU and sky fibers are labeled.}
\label{fig:frame_front}
\end{figure}

The basic design of the Binospec IFU assembly is shown in Figure~\ref{fig:frame}. 
Science target light enters at the location labeled ``Science fiber pickoff mirror", and passes through the magnifier optics and lenslet array. These are bonded to the fibers on the back side, shown in orange in Figure~\ref{fig:frame} and in the photo in Figure~\ref{fig:fibers}. Fibers are then routed in bundles of 20 to one of the 90-degree prisms arranged linearly along the length of the mask frame to form the pseudo slit (32 science plus 4 sky on each side). Off-target light for sky background subtraction enters at its own pickoff mirror, but passes through without additional focusing optics. Sky fibers are randomly divided into bundles of 8, which are evenly dispersed in between the science bundles on the output side. Figures \ref{fig:frame_back} and \ref{fig:frame_front} show the final assembled unit, back and front, respectively.

\begin{figure}[ht!]
\includegraphics[width=\hsize]{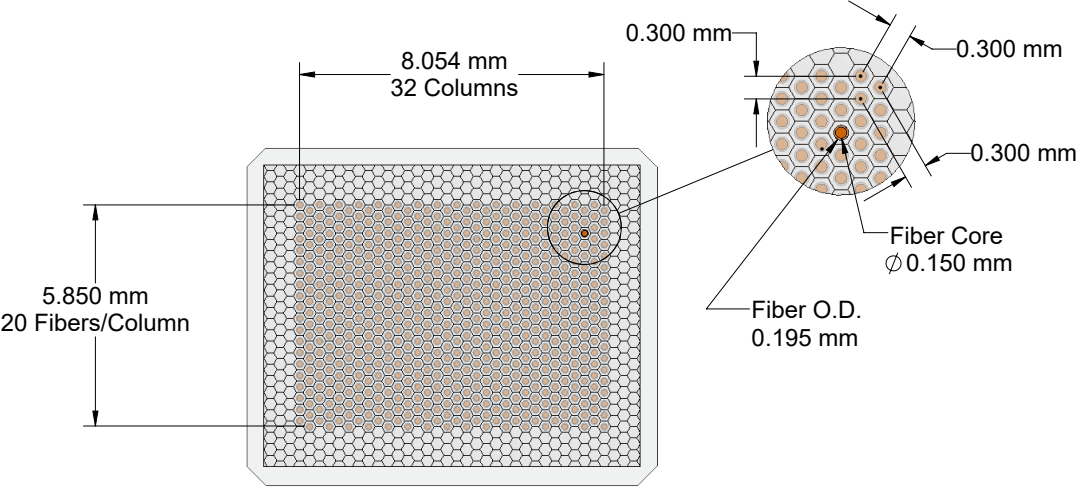}
\caption{Lenslet and fiber layout. The 640 active lenslets are arranged in a 20 by 32 grid. Inset shows fiber diameters relative to lenslet dimensions and spacing.}
\label{fig:lenslet}
\end{figure}

\subsection{Magnifier Optics} \label{sec:optics}
\begin{figure}
\includegraphics[width=\hsize]{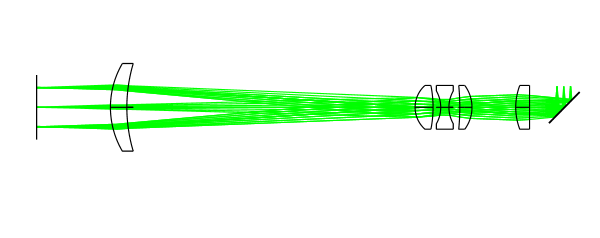}
\caption{IFU magnifier optics.  Light from the MMT at f/5 enters on the right side of the image and is relayed to a 3X magnified focus on the left. The path from focus to focus is 100mm.}
\label{fig:optics}
\end{figure}

We chose an IFU sampling of 0.6$^{\prime\prime}$ to critically sample the mean MMT seeing of 1.2$^{\prime\prime}$, corresponding to a physical scale of 100$\mu$m at the f/5 focus.  The fibers are registered to the hexagonal lenslet array with a precision bored fused silica plate (Figure~\ref{fig:lenslet}).  The plate requires an interfiber wall thickness of $\sim$100$\mu$m, so operating at the plate scale of the f/5 focus leaves no room for an optical fiber.  Magnifier optics are clearly required: 2X magnification leaves 100$\mu$m for the fiber and 3X magnification (adopted) leaves 200$\mu$m for the fiber.  We designed a five-lens system to relay a patch of the f/5.3 MMT field to f/15.9 to illuminate the lenslet array (Figure~\ref{fig:optics}) and Table~\ref{opticstable}).  The magnifier optics were manufactured by Optics Technology in Pittsford, NY and supplied with broadband anti-reflection coatings.  The lenslet array was fabricated and anti-reflection coated by Advanced Microoptic Systems with a pitch of 300$\mu$m, a thickness of 3.3mm and a focal length of 2.3mm.  The best spatial resolution is attained with a magnifier focus set 2.3 mm into the lenslet array. Each lenslet element forms a pupil image 149$\mu$m in diameter, on the rear, planar surface of the array, well matched to the 150$\mu$m fiber. A raytrace shows that 99\% of the light enters the fiber at an incidence angle $<$5.39$^\circ$, corresponding the the f/5.3 input focal ratio expected by Binospec.

\begin{deluxetable}{lrrrr}
\tablecaption{IFU Magnifier Optics\label{opticstable}}
\tablewidth{700pt}
\tabletypesize{\scriptsize}
\tablehead{
\colhead{Name} & \colhead{Radius} & 
\colhead{Thickness} & \colhead{Material} & \colhead{Size(mm)}
} 
\startdata
Focal Surface  & Infinity &  3.689  	&       & 3.18 x 2.53\\
Fold Mirror    & Infinity &  6.311   &  MIRROR	& 8\\
Lens 1 (front) & Infinity &  2.524   & S-FPM2   & 8\\
Lens 1 (back)  & -11.215  &  7.990   &          & 8\\
Lens 2 (front) &   6.730  &  2.223   & S-FPL51  & 8\\
Lens 2 (back)  &  41.255  &  1.989   &          & 8\\
Lens 3 (front) &  -6.062  &  1.498   & S-TIL27  & 8\\
Lens 3 (back)  &   6.062  &  1.397   &          & 8\\
Lens 4 (front) &  20.386  &  3.283   & S-FPL51  & 8\\
Lens 4 (back)  &  -5.338  & 52.643   &          & 8\\
Lens 5 (front) & -28.164  &  3.000   & S-FPM3   & 16 x 10.4\\
Lens 5 (back)  & -15.802  & 13.481   &          & 16 x 10.4 \\
Focus          & Infinity &  -       &          & 9.29 x 7.34 \\
\enddata	
\end{deluxetable}

\section{Testing and Characterization}
\begin{figure}[ht!]
    \centering
    \includegraphics[width=\linewidth]{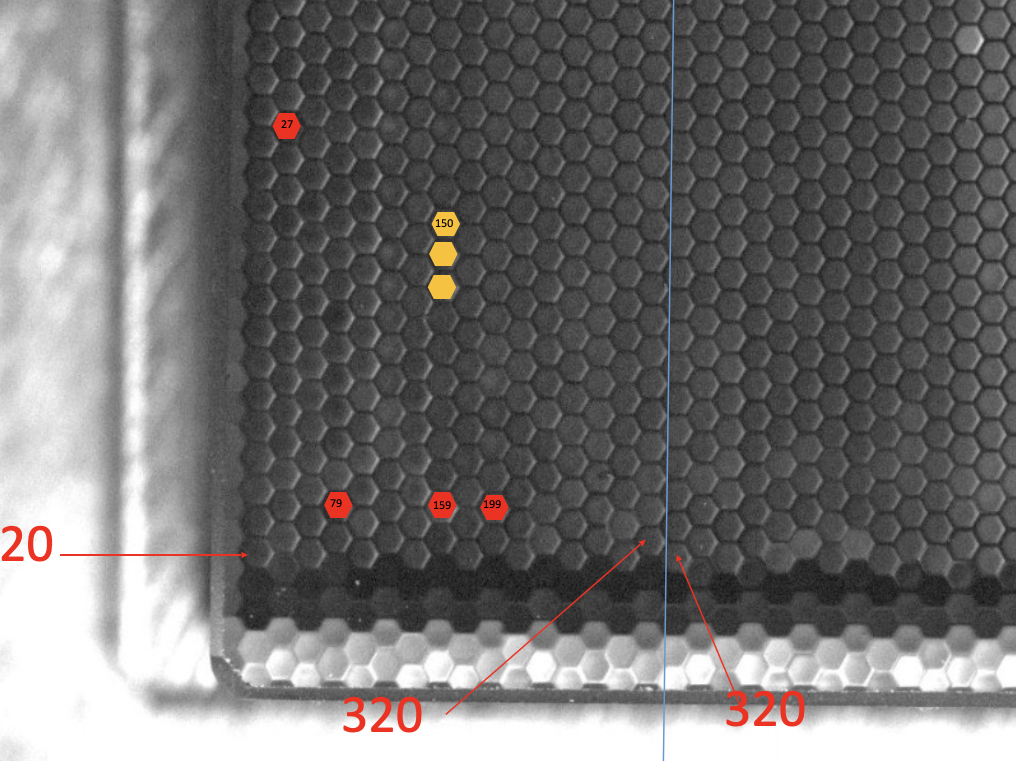}
    \caption{Image showing fibers identified as dead from lab testing (in red), and those that were swapped on the science CCD side (in orange). The light blue line shows the divide between lenslets routed to one channel of Binospec and those routed to the other channel. Red numbers show the numbering convention of the fibers, starting at the outer top corner of each side, and ending at fiber 320 at the bottom center (designated A1-A320 and B1-B320 for the two sides).}
    \label{fig:ifu_fibers}
\end{figure}

After all fibers were routed from the lenslet array to the fiber block prisms, but before installing covers, the light path through the fibers was tested. A light source was scanned along the back spectrograph-facing slits of the assembly, and a video camera was set up to record the light emerging on the science lenslet side. This allowed us to verify and record the mapping between lenslet position in the focal plane and the fiber output position on the CCD side. This mapping is essential for proper reconstruction of a data cube from the reduced data of the individual fibers. We identified 4 dead fibers which passed no light, and found 3 fibers which were swapped from their expected ordering on the CCD side. (See Figure~\ref{fig:ifu_fibers}).

After final assembly and shipment to the telescope, the fit of the IFU into Binospec was carefully tested, first by manually sliding the unit into an open slot in the Binospec mask elevator to verify clearances. After slight modifications, we verified that Binospec's mask arm could grab the IFU unit and precisely slide it into place. Once that mechanical operation was verified, we were ready to proceed with science commissioning.

\section{Observation Planning and Operation Software}
We created a dedicated version of the Binospec mask design software, BinoMask, to help potential users visualize the IFU placement over their target of interest, and verify suitable guide stars, WFS stars, and placement of the sky fiber bundle.\footnote{\url{https://scheduler.mmto.arizona.edu/BinoMask/index.php?ifu=true}}

At the telescope, queue observers align the IFU using a technique similar to that used for long slit or mask observations. First, the telescope operator slews to the IFU target. Then, the observer executes a command that offsets the telescope to place the target within the IFU field of view. Expected guide star positions are calculated for the offset position, and then final alignment is achieved by applying pointing offsets to place the detected guide stars at the pre-calculated coordinates. Thanks to Binospec's stiff and well-calibrated guider mechanism and optics, this procedure achieves alignment better than 0.1$^{\prime\prime}$, far better than required for the IFU.

The dimensions of the guide star cutout regions in the IFU mask frame are stored in the Binospec database \citep{2019PASP..131g5005K}. These dimensions are slightly different than those used in Longslit or Imaging mode, and so we must store and use the proper boundaries when selecting available guide stars for any given instrument pointing.

The {\sc bobserve} instrument control software has special templates for IFU observations, which require somewhat longer exposure times for calibration frames. Also, because of the offset of the science IFU block with respect to the center of the mask assembly, there is a shift of the central wavelength compared to long-slit and MOS observations, which is taken into account while selecting a suitable arc line for the active flexure control system. This shift varies with the selected grating and nominal central wavelength. For PIs and observers this is completely transparent but managed automatically by the {\sc OnDeck} staging and the automatic queue of {\sc bobserve} allowing to take series of images and calibrations if necessary. 

Currently, the guider software allows for dithering offsets up-to $\pm$7.5~arcsec in both directions and enables us to observe small mosaics up-to 2$\times$2 FoVs (30$\times$24~arcsec on the sky) without re-acquiring guide stars. The dithering pattern and amount is set by the observer in {\sc OnDeck} or the automatic queue. In the future, we plan on extending the mosaicing mode to handle larger offsets along the long side of the IFU FoV, which is aligned with the direction of guide star boxes in the IFU mask assembly. In principle, we should be able cover the sky area of up-to 240$\times$24~arcsec without re-acquiring guide stars to map extended targets (e.g. edge-on galaxies or long filaments in H{\sc ii} regions or supernova remnants.

\section{Commissioning and Science Verification.} \label{sec:commsciv}

\begin{figure*}
    \centering
    \includegraphics[width=\hsize]{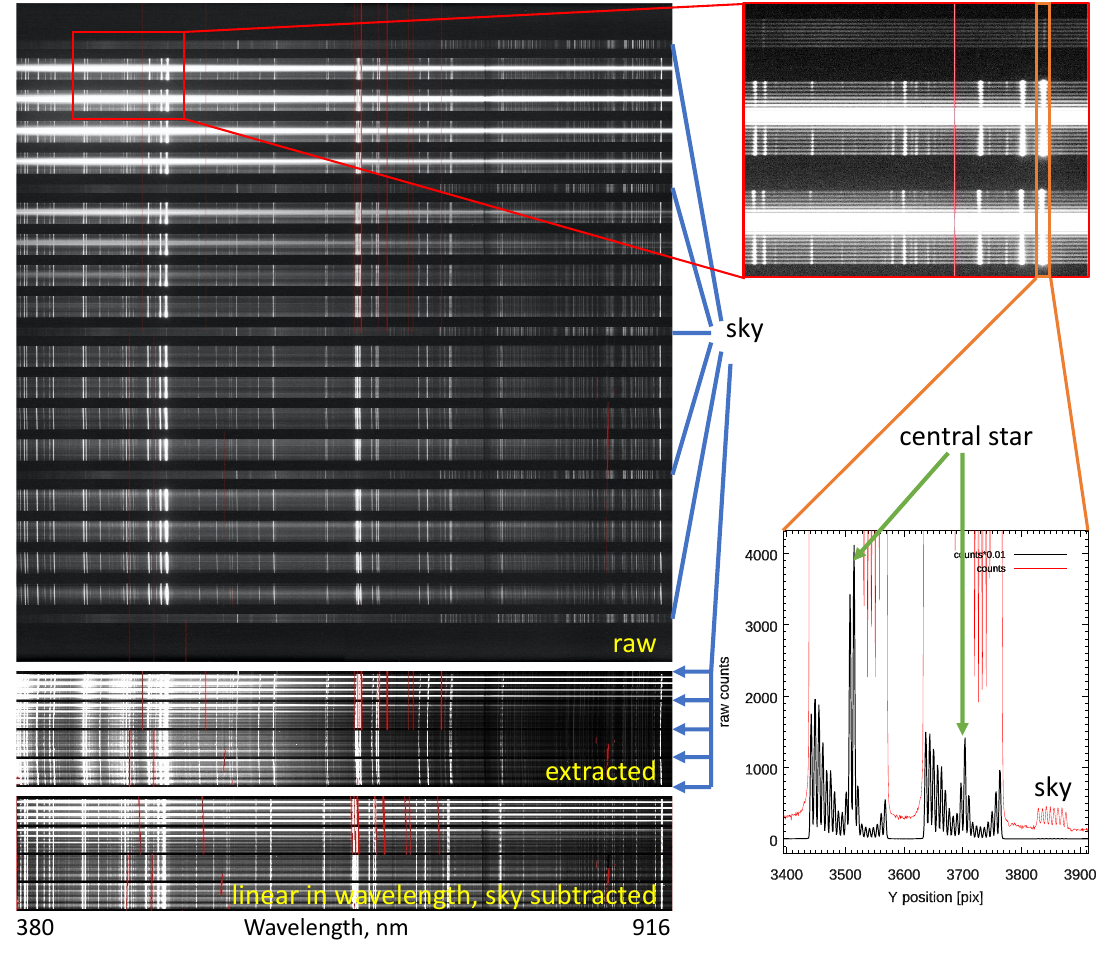}
    \caption{Left (top to bottom): A full raw image (Side A) of the commissioning target \object{NGC~2392}, extracted spectra, extracted spectra linearized in wavelength with night sky background subtracted. Top right panel: a zoomed fragment of the raw image showing one block of sky fibers (above) and two blocks of science fibers (below). Bottom right panel: a cut through the image across the fibers at the position of the [O{\sc iii}] 5007~\AA\ line; fiber traces, which correspond to the central star and sky background (enhanced by a factor of 100) are identified.}
    \label{fig:raw_cutout}
\end{figure*}

The commissioning and science verification of Binospec-IFU was performed in December 2023.  The commissioning tasks included:
\begin{itemize}
    \item \emph{Confirm the focus position and the calibration of central wavelength versus grating rotation.} We obtained arc and internal flat field spectra and found that there is no significant focus offset between the focus for IFU fibers and MOS aperture masks.  The central wavelength calibration offsets agree with the Binospec optical model predictions within two pixels.
    
    \item \emph{Confirm the field orientation and offsets of the IFU relative to the guiders and measure the IFU angular scale.} We observed a four-star asterism at low Galactic latitude and confirmed the spatial scale to be within 1\%\ of 0.6$\asec$/lenslet. The location of the science IFU FoV with respect to the guider probes was established to better than 0.1$\asec$. 
    
    \item \emph{Determine the IFU fiber illumination correction and line-spread function.} We obtained day-time sky flats using sunlight entering through the partially open rear MMT dome shutters. We compared the sky and internal flats and found that the illumination corrections for all three Binospec gratings and central wavelengths are consistent.  From the same sky flats we determined the spectral line spread function as a function of wavelength and spatial position by fitting a high-resolution Solar spectrum to spectra in each fiber. We use a Gauss-Hermite parametrization with 3-rd and 4-th order Hermite polynomials $h3$ and $h4$.
    
    \item \emph{Test the data reduction and image reconstruction algorithms.} We observed \object{NGC~2392}, a planetary nebula with a bright central star and compared reconstructed images in several optical emission lines (O{\sc iii}, H$\alpha$ and others) to smoothed narrow-band images obtained with the Hubble Space Telescope. A cutout from the raw image of \object{NGC-2392} is shown in Figure~\ref{fig:raw_cutout}. The image shows the individual fiber traces for two 20-fiber science blocks, as well as a single 8-fiber sky block. Nebular emission lines are clearly visible in the science fibers, distinct from the sky emission lines visible in the sky block.
\end{itemize}

Science verification targets were added to the Binospec queue, interspersed with long-slit, MOS, and imaging observations. The main goals of science verification were:
\begin{itemize}
    \item Measure the Binospec-IFU throughput in a wide range of grating/central wavelength combinations. We observed several spectrophotometric standard stars in photometric conditions, and compared the observed count rates to absolute spectrophotometry in the literature.
    
    \item Assess the instrument performance for bright emission-line targets by observing two Seyfert galaxies with radio jets.
\end{itemize}

\section{Data Reduction Software} \label{sec:software}

Data reduction recipes for the Binospec IFU are integrated into the official Binospec data reduction pipeline \citep{2019PASP..131g5005K} maintained by the Telescope Data Center at the Center for Astrophysics (\href{https://bitbucket.org/chil\_sai/binospec/wiki/Home}{https://bitbucket.org/chil\_sai/binospec/wiki/Home}). Some of the IFU-specific algorithms in the pipeline are based on those from the IFU data reduction package developed for the Multi-Pupil Fiber Spectrograph (MPFS) at the Russian 6m telescope described in  \citet{2007MNRAS.376.1033C}. This package was enhanced to reduce VIMOS-IFU \citep{2009ApJ...697L.111C} and Potsdam Multi-Aperture Spectrograph \citep{2010MNRAS.405L..11C} data. For the Binospec IFU we modified the spectral tracing and extraction routines. The pipeline computes and propagates variance frames from photon statistics to provide realistic error estimates.  We determine relative illumination corrections and matched spectral line spread functions for the sky IFU sky fibers (also used in the GMOS-IFU \citep{2002PASP..114..892A}, MPFS \citep{2001sdcm.conf..103A}, and SCORPIO2-IFU \citep{2018AstBu..73..373A}) 

Overall, the data reduction steps in the IFU pipeline are similar to those in the multiple slitlet reduction excepting that IFU flat fielding is performed after spectral extraction, not before as with slitlets.  The reduction steps are:

\begin{itemize}
\item CCD overscan removal: we apply a non-linearity correction and correct for the different gains of the four amplifiers in each Binospec CCD. 

\item Geometrical distortion mapping:  in each Binospec channel we cross-correlate the spatial profile in seven segments at 32 wavelength positions with a reference profile extracted at the central wavelength. We use a 3$\times$3 order 2D-polynomial to fit the distortion map. 

\item Fiber profile tracing and scattered light subtraction: this step is carried out in two iterations. First, the fiber traces are identified using a peak detection algorithm, and then their shapes and positions are fit by 3rd to 6th order Gauss-Hermite functions plus an additive offset. We start by fitting the first fiber in each sky or science prism block and subtracting the best-fitting function from the flat field profile. Then we fit the subsequent fiber and also subtract it from the profile, then refit the current fiber in a wider window. This approach allows us to better handle cross-talk between neighboring fiber traces, which overlap at 5--7\%\ of the peak intensity. From the result of the first iteration of tracing we determine the gaps between fiber blocks and build the scattered light model, which we then subtract from all 2D frames (science and calibrations). Then we proceed to the second iteration which is identical to the first one but the Gauss-Hermite functions now are fitted without the additive term. Here we also assume a smooth behavior of Hermite coefficients as a function of both wavelength and position on the pseudo-slit and fix them in the fitting procedure leaving free only width and central position of each trace peak.

\item Spectral extraction: we use scattered light subtracted frames and the normalized Gauss-Hermite profiles of spectral traces obtained at the previous step to extract the signal at each position along the wavelength using optimal extraction, i.e. applying the weights that correspond to the profile shape. We do not extract individual fibers within an extraction window as is commonly done. Instead, we extract the signal from all the fibers within a fiber block of 8 (sky) or 20 (science) fibers using a constrained linear inversion to force the positivity of an extracted signal from each fiber. This step is implemented by reducing a constrained linear fit to a positively defined quadratic programming problem. We use \emph{NaN} (not a number) to mask the flux values in case a given trace has more than 70\%\ of pixels affected by cosmic ray hits or CCD imperfections. Our approach automatically handles the crosstalk between neighboring fibers assuming that we derived accurate trace profiles at the previous data reduction step. Through the rest of the pipeline reduction, the sky and science blocks of extracted spectra are treated very similarly to MOS slitlets but without applying a distortion correction because the extracted stacked spectra are rectified by construction.

\item Flat field correction: we use an internal incandescent lamp + LED flat field source for field illumination and vignetting correction. Pixel-to-pixel sensitivity variations are negligible after the extraction that averages five pixels at each wavelength position. We normalize the flat field shape to that derived in the central part of the SN1 Binospec channel to cross-calibrate the two Binospec channels. We apply the sky flat illumination correction.

\item Wavelength calibration: we treat each fiber block as a MOS slitlet in the original Binospec pipeline. We use the Binospec optical model to calculate an initial wavelength solution accurate to two CCD pixels. We then identify Ar lines in calibration exposures or airglow lines in the observed spectrum (where possible) or a combination of the two. We fit a preliminary wavelength solution using a low-order two-dimensional polynomial (Order 3 to 4 in the wavelength direction and order 2 to 3 in the spatial direction).  We slightly adjust the zero-point of the wavelength solution in each fiber within a block by cross correlating a linearized calibration spectrum with a reference spectrum from the center of the pseudo-slit. This procedure accounts for the small fiber offsets of individual fibers.  We typically achieve a wavelength solution better than 1/20th of a pixel RMS.

\item Secondary illumination correction: we measure fluxes in one or more airglow lines (e.g. [O{\sc i}] $\lambda=5577$~\AA) by fitting the lines with Gaussian-Hermite profiles and computing a secondary fiber-to-fiber illumination correction assuming that the airglow line flux is constant across the IFU field of view.  When several lines are used, we can correct wavelength dependent corrections.  However, we find that these corrections are $<$1.5\%.  We compute the barycentric correction and apply it to the wavelength solution.

\item Sky background subtraction: we use a variant of the sky subtraction technique proposed by \citet{2003PASP..115..688K} modified for MOS and fiber-lens IFU data reduction to use a global sky background model as described in \citet{2015PASP..127..406C}. The sky model is computed with extracted spectra prior to the wavelength linearization. We use a $b$-spline parametrization of an oversampled sky spectrum assembled from all sky background fibers and evaluate the model at each wavelength and pseudo-slit position. The low-order Legendre polynomial terms are added to the $b$-spline parametrization along the pseudo-slit to handle smooth variations of the sky background originating from line-spread-function variations and residual flat-fielding imperfections. This approach allows one to eliminate an extra re-sampling step and, therefore, yields much better emission line sky subtraction.  We apply a partial parametric deconvolution to the sky model to account for the $\sim$10\%\ FWHM LSF difference between sky background and science fibers. This is done by convolving the signal with the kernel in the form: $[-a, 1+a+b, b]; 0 \leq a,b \ll 1$ where $a$ and $b$ are of the order of 0.05--0.1 (the values were determined empirically for each spectral setup).  We reach sky subtraction close to the Poisson statistical limit.

\item Flux calibration and telluric correction: If a spectroscopic standard star is observed,  the pipeline performs absolute spectrophotometric calibration and calculates the instrument throughput. By comparing our pipeline products against flux calibrated spectra from large surveys such as Sloan Digital Sky Survey, we estimate that our absolute flux calibration quality is $\sim$3\%\ in photometric conditions when the standard star is within 30~deg of the target.  Any {\sc a0v} non-variable star with low Galactic foreground extinction can be used for less precise flux calibration.  We can run a postprocessing code to correct for telluric absorptions.  We fit a grid of atmospheric transmission models at the MMT elevation with a star or galaxy spectrum. The approach is described in \citet{2023ApJS..266...11B} and corrects better than 2\%\ in spectral regions with $<50$\%\ telluric absorption (the entire spectral range at Binospec resolution).
\end{itemize}

The final data products are presented in the {\sc fits} format either in the form of linearized flux-calibrated stacked spectra with a table for individual fibers or as a regularly gridded oversampled 3-dimensional data cube. The {\sc fits} World Coordinate System (WCS) keywords are provided for the calibration of both wavelength and sky coordinates.  We encourage scientific users to use stacked spectra for scientific analysis while the data cubes are useful for data quality assessment in quick look tools, e.g. {\sc SAOImage ds9} \citep{2003ASPC..295..489J}. 

\section{Instrument performance} \label{sec:performance}

\subsection{Throughput measurements}

\begin{figure}
\includegraphics[width=\hsize]{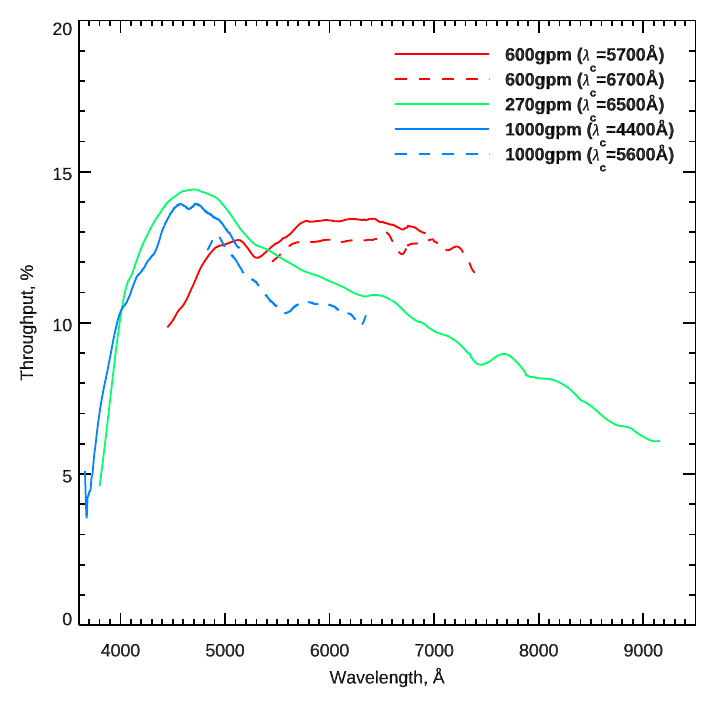}
\caption{Binospec IFU throughput (including the telescope and excluding the atmosphere) in the center of the field of view with different gratings and central wavelengths derived from observations of the star \object{G191-B2B} collected on the night of 2023/Dec/13. \label{fig_binoifu_throughput}}
\end{figure}

We measured the throughput of the Binospec IFU in different setups using the spectrophotometric standard star \object{G191-B2B} during the second night of the commissioning run. This star is a very hot white dwarf \citep{1995AJ....110.1316B}, therefore there is a significant second order contamination at wavelengths $\lambda>7500$~\AA\ even with the \textit{LP3800} blocking filter. The throughput of the system peaks at about 13.5--14.5\%\ (atmosphere excluded using standard extinction correction for Mt.~Hopkins, telescope included). The IFU throughput curves from the center of the field of view (see Fig.~\ref{fig_binoifu_throughput}) repeat those measured in the longslit mode \citep{2019PASP..131g5004F} with about 50\%\ lower efficiency, which we attribute primarily to optical fiber focal ratio degradation with a small contribution from the IFU relay optics and fold mirror.

The throughput, assesed with twilight sky observations, varies across the field of view. We see a 25\% throughput drop for two rows of microlenses at the edge of the field of view in Side A. In the rest of the IFU the throughput variations stay within $\pm6$\% of the mean value at the center of the field. These variations translate to small differences in the signal-to-noise ratio across the field but do not affect the quality of absolute flux calibration because they are accounted for by the data reduction pipeline during a secondary flat field correction using sky airglow lines.

\subsection{Spectral resolution}

We measured spectral resolution and its variations across the field of view and wavelength range by fitting a high-resolution Solar spectrum against high signal-to-noise twilight flats using a Gauss-Hermite representation of the spectral line-spread-function (LSF). Variation of the LSF width across the field of view is negligible ($\sigma(R)\approx2$\%). Overall, the spectral resolving power delivered by the IFU is around 25~\%\ higher than that in the longslit mode when using a 1~arcsec-wide slit. For example, in the high-resolution (1000gpm grating) blue setup it monotonically grows from $R=4100$ ($\sigma_{\mathrm{inst}}=30$~km~s$^{-1}$) at $\lambda=3850$~\AA\ to $R=7000$ ($\sigma_{\mathrm{inst}}=18$~km~s$^{-1}$)  at $\lambda=5300$~\AA. The same applies to other setups, e.g. the 600gpm grating delivers the spectral resolving power $R=5,300$ (($\sigma_{\mathrm{inst}}=24$~km~s$^{-1}$) around H$\alpha$ ($\lambda=6563$\AA) compared to $R=4,200$ in the longslit/MOS mode with a 1~arcsec-wide slit.

There is a modest deviation of the LSF from a pure Gaussian manifested by the coefficient $h_4\approx-0.05$ (slightly $\Pi$-shaped profile). At the same time, $h_3$, which manifests the LSF asymmetry, is consistent with 0 in all setups. 

\subsection{Comparison of Binospec IFU with other IFU spectrographs}

The performance of Binospec IFU compares quite well to IFU spectrographs operated at intermediate-sized and large telescopes.  The Binospec IFU spectral resolution with the 1000gpm and 600gpm gratings exceeds that of most other instruments except the Keck Cosmic Web Imager \citep[KCWI;][]{2018ApJ...864...93M} with medium and high-resolution gratings and medium/small slicers operated at the 10-m Keck telescope and FLAMES in the ARGUS mode operated at the 8-m ESO VLT. 

Our low-resolution 270gpm setup compares quite well to the B600 grating of GMOS-N/S IFU at the twin 8-m Gemini telescopes and a 600~gpm grating in the lens array mode of the Potsdam Multi-Aperture Spectrograph \citep[PMAS;][]{2006SPIE.6269E..0HR} operated at the 3.5-m telescope of the Calar-Alto Observatory in Spain and the IFU mode of the SCORPIO universal spectrograph \citep{2018AstBu..73..373A} at the Russian 6-m telescope. At the same time, Binospec IFU provides twice as broad wavelength coverage at the same spectral resolution.

The high-resolution blue setup with the 1000gpm grating has a throughput $\sim$25\% lower than that of the most blue-sensitive IFU spectrograph available at a large telescope KCWI\footnote{\url{https://www2.keck.hawaii.edu/inst/kcwi/thoughput.html}}.  KCWI uses an image slicer rather than fibers.  With the medium blue grating and medium slicer KCWI offers comparable spectral resolution. However, the Binospec IFU covers almost twice the wavelength range (1550~\AA\ vs 850~\AA). The low-resolution Binospec IFU configuration has about 50\% the throughput of KCWI (low-resolution grating, medium slicer) but provides twice the wavelength coverage at the same spectral resolution.

The Binospec IFU throughput is nearly twice as high as some other fiber/lenslet IFUs, such as PMAS (our own measurement), and SCORPIO-IFU \citep{2018AstBu..73..373A} with a comparable field-of-view, plus the Binospec IFU has much broader wavelength coverage.  We calculate the throughput of the GMOS-N/S IFU by combining the throughput of the IFU block \citep{2002PASP..114..892A} with the longslit throughput\footnote{\url{https://www.gemini.edu/instrumentation/gmos/capability\#Sensitivity}}.  The Binospec IFU throughput is about 50\% higher than this combination.

\section{Examples of Science Verification Observations} \label{sec:science}

\subsection{NGC~2392, a bright planetary nebula}

\begin{figure*}
    \centering
    \includegraphics[trim=3mm 2mm 5mm 3mm,clip,width=0.24\hsize]{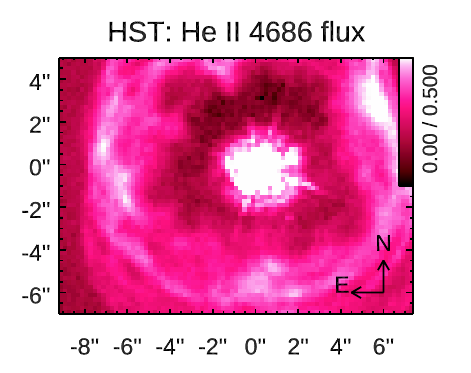}
    \includegraphics[trim=3mm 2mm 5mm 3mm,clip,width=0.24\hsize]{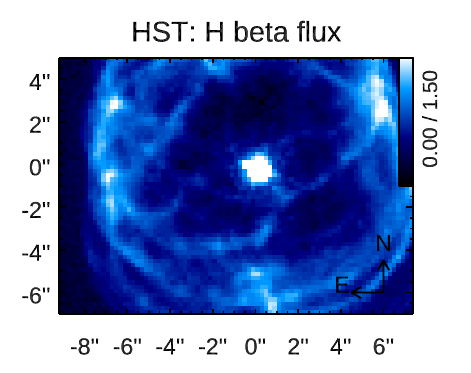}
    \includegraphics[trim=3mm 2mm 5mm 3mm,clip,width=0.24\hsize]{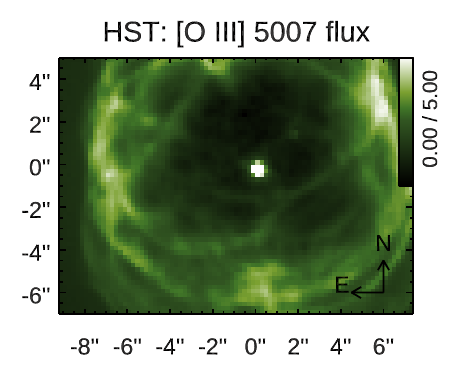}
    \includegraphics[trim=3mm 2mm 5mm 3mm,clip,width=0.24\hsize]{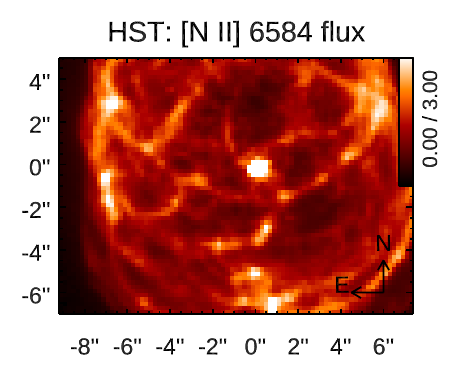}
    \includegraphics[trim=3mm 2mm 5mm 3mm,clip,width=0.24\hsize]{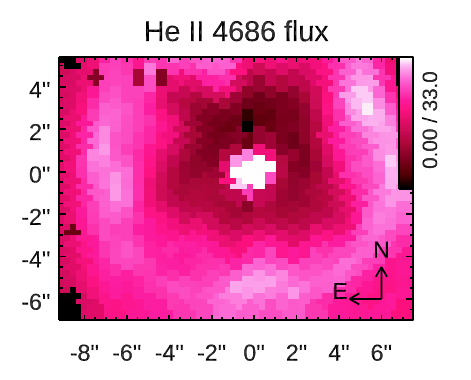}
    \includegraphics[trim=3mm 2mm 5mm 3mm,clip,width=0.24\hsize]{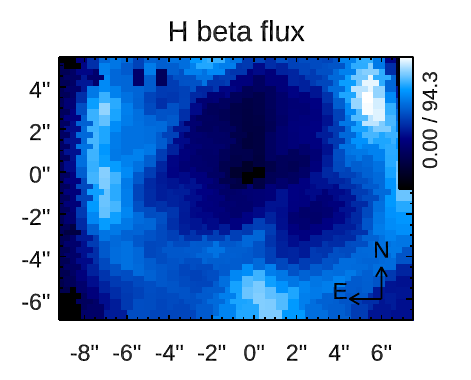}
    \includegraphics[trim=3mm 2mm 5mm 3mm,clip,width=0.24\hsize]{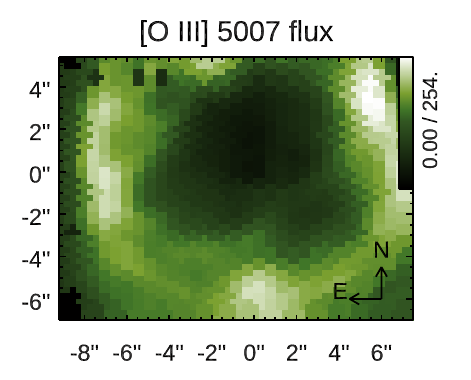}
    \includegraphics[trim=3mm 2mm 5mm 3mm,clip,width=0.24\hsize]{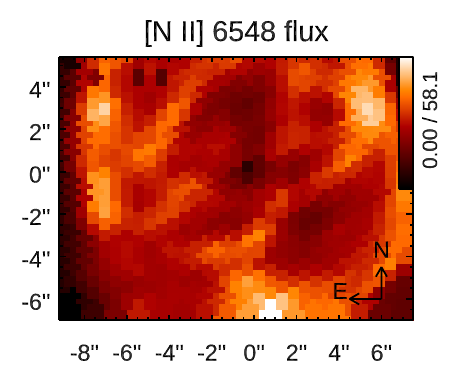}
    \includegraphics[trim=3mm 2mm 5mm 3mm,clip,width=0.24\hsize]{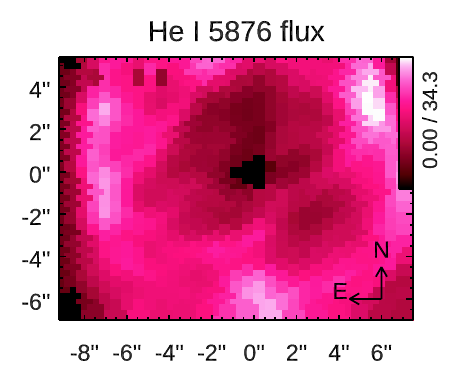}
    \includegraphics[trim=3mm 2mm 5mm 3mm,clip,width=0.24\hsize]{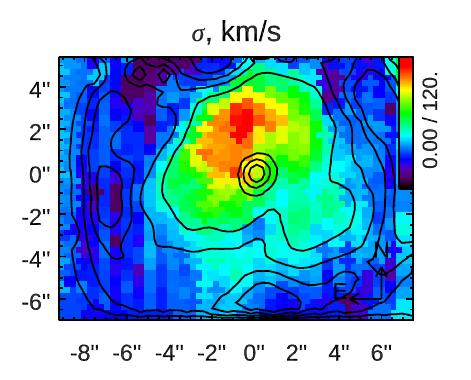}
    \includegraphics[trim=3mm 2mm 5mm 3mm,clip,width=0.24\hsize]{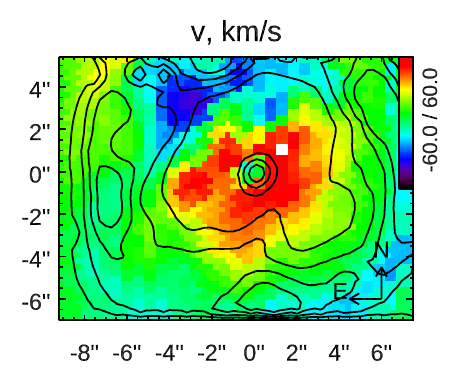}
    \includegraphics[trim=3mm 2mm 5mm 3mm,clip,width=0.24\hsize]{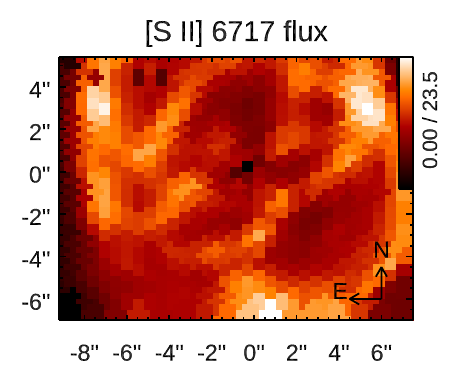}\\
    \vskip 2mm
    \includegraphics[trim=6mm 2mm 5mm 10mm,clip,width=\hsize]{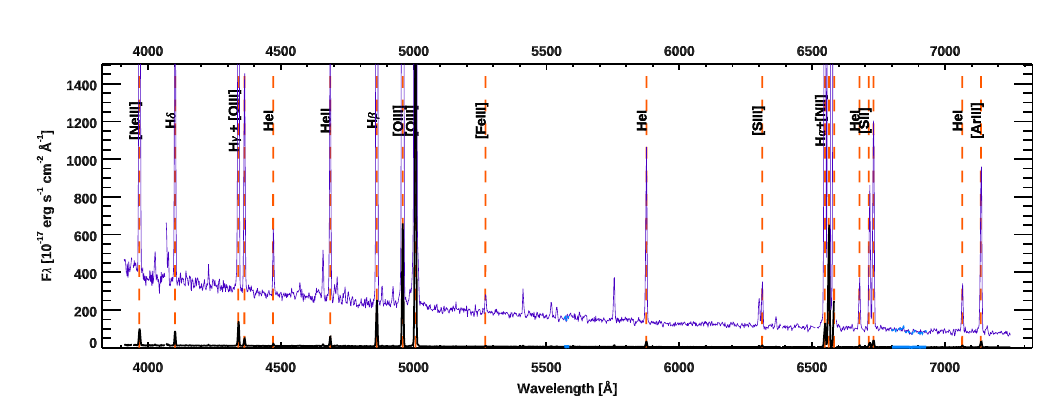}
    \caption{Top row: Hubble Space Telescope narrow-band images of the planetary nebula NGC~3292 in 4 emission lines, He{\sc ii}, H$\beta$, [O{\sc iii}], [N{\sc ii}].  Second row: images reconstructed from the Binospec data cube in the same emission features. A different [N{\sc ii}] line ($\lambda = 6548$~\AA) was used to avoid CCD artifacts.  Third row: two additional Binospec line maps (He{\sc i} and [S{\sc ii}]) and maps of velocity dispersion and radial velocity. We use square root flux scaling to increase the displayed dynamic range. The black contours in the kinematics fields were computed from the [O{\sc iii}] line flux. Bottom panel: an extracted spectrum from a single lenslet (black solid line) with emission lines identified. The same spectrum scaled 30X is shown in purple to enhance faint emission lines.}
    \label{fig_ngc2392}
\end{figure*}

To demonstrate the IFU performance on targets with high dynamic range, we observed NGC~2392, the Eskimo nebula. NGC~2392 has a 10.5~mag central hot star and over a hundred identified emission lines.  NGC~2392 was observed during the first commissioning night, 2023/Dec/12, through thin clouds with 1.3~arcsec seeing.  We used the low-resolution 270~gpm gratings centered at 6,500~\AA\ with the $LP3800$ blocking filter. This configuration covers 3800 to 9300~\AA\ at $R\sim2200$. 

In Fig.~\ref{fig_ngc2392} we demonstrate the results of the full spectrum fitting of a Binospec IFU datacube for NGC~2392. We used an updated version of the {\sc NBursts} full spectrum fitting technique \citep{2007IAUS..241..175C,2007MNRAS.376.1033C} with a list of emission lines from a recent Euclid planetary nebula spectral atlas \citep{2023A&A...674A.172E} and a template spectrum for the central star collected with the Hubble Space Telescope \citep{2018A&A...615A.115K}. Fig.~\ref{fig_ngc2392} (second and third rows) shows radial velocity and velocity dispersion maps of the ionized gas derived from emission lines fitted by a single-component Gauss-Hermite function and reconstructed flux maps in several strong emission lines.

In Fig.~\ref{fig_ngc2392} we qualitatively compare images in the He{\sc ii}, H$\beta$, [O{\sc iii}], and [N{\sc ii}] lines reconstructed from the IFU datacube (second row) with narrow-line images of the same source collected with the Hubble Space Telescope extracted in the same region (top row). Because the central star has significant emission in the He{\sc ii} line it is visible visible in the corresponding Binospec IFU image.  In the remaining images its contribution is subtracted off by the fitting procedure.  The HST narrow-band images show the central star in each image because no continuum has been removed. The bottom row of Fig.~\ref{fig_ngc2392} displays an extracted spectrum of a bright region of the nebula.

\subsection{GMP~2640, a low-luminosity post-starburst galaxy}

\begin{figure*}
    \centering
    \includegraphics[trim=3mm 2mm 5mm 3mm,clip,width=0.24\hsize]{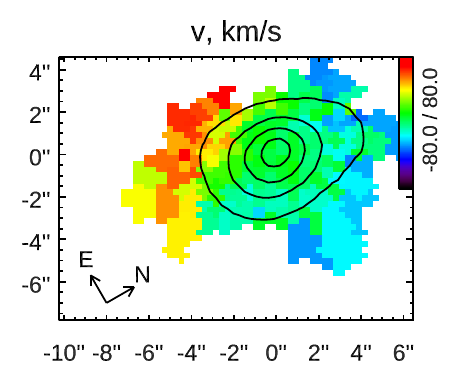}
    \includegraphics[trim=3mm 2mm 5mm 3mm,clip,width=0.24\hsize]{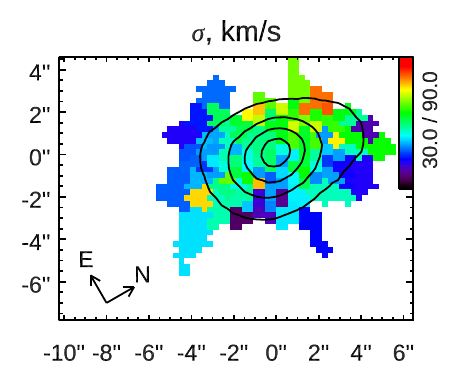}
    \includegraphics[trim=3mm 2mm 5mm 3mm,clip,width=0.24\hsize]{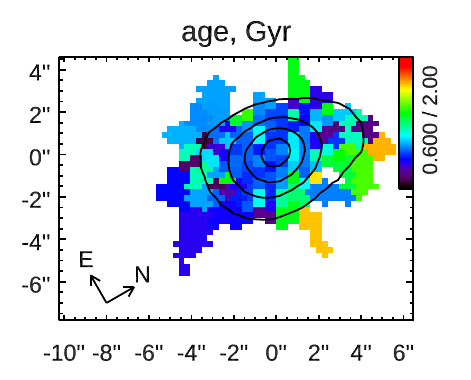}
    \includegraphics[trim=3mm 2mm 5mm 3mm,clip,width=0.24\hsize]{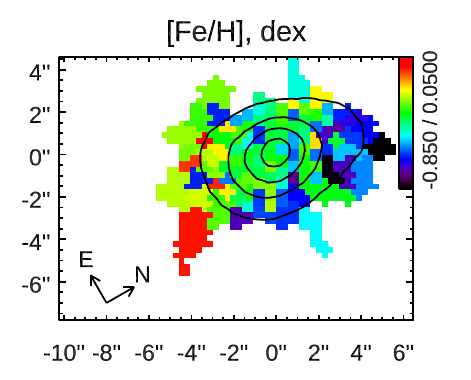}
    \includegraphics[trim=8mm 2mm 5mm 6mm,clip,width=\hsize]{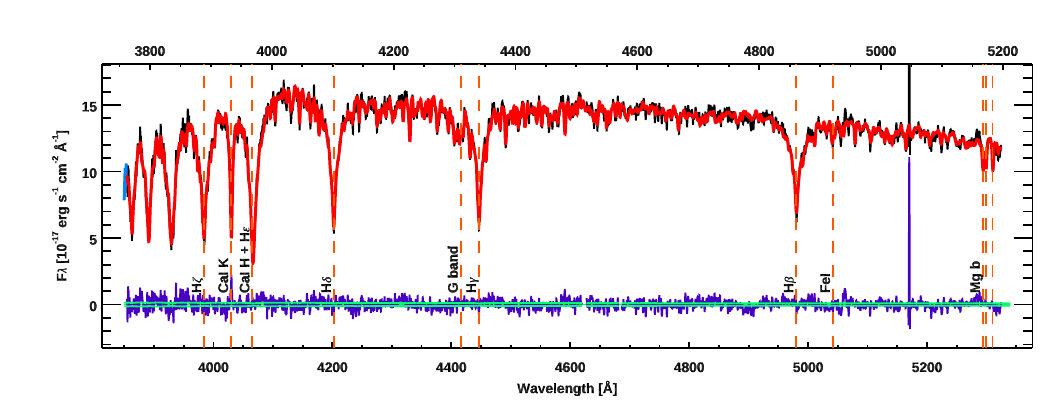}
    \caption{Top row left to right: maps of stellar radial velocity, velocity dispersion, mean stellar ages and metallicities in a low-mass, diffuse, post-starburst galaxy GMP~2640 in the Coma cluster. Bottom: a coadded spectrum of the 7 central lenslets (black) with main spectral features identified and flux uncertainties (green lines around zero), the best fitting stellar population model (red), and fitting residuals (purple). Observed and rest-frame wavelengths are shown on the bottom and top axes correspondingly.  Black contours were reconstructed from the continuum flux at rest-frame wavelengths from 4000--5000~\AA.}
    \label{fig_gmp2640}
\end{figure*}

To assess the performance of the instrument on low-surface brightness absorption-line targets in the intermediate-resolution blue setup with the 1000~gpm grating centered at 4,500~\AA, we observed GMP~2640, a low-mass ($M_*=2.4\cdot10^9 M_{\odot}$) post-starburst galaxy in the Coma cluster, which had been previously observed with Binospec with the same grating with a slit mask. \citep{2021NatAs...5.1308G}. The total integration time was 2~h, with 40~min of integration on 2023/Dec/17, and the remaining 1:20~min on 2024/Jan/13. Both datasets were acquired under good transparency, dark moonless sky, and average seeing conditions ($\sim1.1-1.2$~arcsec FWHM). We observed the spectrophotometric star GD~153 to calibrate each science observation.

In Fig.~\ref{fig_gmp2640} we present 2D maps of of stellar kinematics (radial velocity and velocity dispersion) and stellar population (mean age and metallicity) of GMP~2640 obtained from the full spectrum fitting of the Binospec-IFU data cube using the {\sc nbursts} full spectrum fitting technique. The data cube was adaptively binned using the Voronoi adaptive binning approach \citep{2003MNRAS.342..345C} to produce a minimum signal-to-noise ratio of five per bin for the radial velocity map and seven for the other three maps. The bottom panel of Fig.~\ref{fig_gmp2640} presents an extracted spectrum from the seven central lenslets of the data cube, the galaxy nucleus and 6 lenslets around it. The 2D map immediately reveals that the major axis of the radial velocity field is rotated $\sim$30$^{\circ}$ with respect to the principal photometric axis, a feature that would be missed by a long-slit aligned with the major axis.

As \citet{2020PASP..132f4503C} demonstrated, internal kinematics statistical uncertainties ($v, \sigma$) measured with full spectrum fitting grow linearly as a function of $\sqrt{\sigma_{\mbox{source}}^2 + \sigma_{\mbox{inst}}^2}$ and drop inversely proportional to $SNR$. If we measure low-velocity dispersion targets with $\sigma_{\mbox{source}} < \sigma_{\mbox{inst}}$ with the Binospec-IFU, the IFU's higher spectral resolution nearly compensates for the lower IFU throughput. Additionally, the velocity field contains a lot more information than a longslit slice, in particular for dynamical modelling of galaxies.  Measuring radial velocities along the minor axis even without velocity dispersion measurements allow us to constrain the stellar orbital anisotropy \citep[see e.g.][]{2023MNRAS.520.6312A}.

\section{Conclusions} \label{sec:concl}

The commissioning of the Binospec IFU significantly expands the utility of this workhorse spectrograph.  Although the total throughput of the integrated IFU is $\sim$50\% lower than a wide Binospec slit ($5\asec$), the throughput is comparable to a 1$\asec$ slit in good seeing and considerably higher in poor seeing. The throughput of the Binospec IFU compares favorably to many operating IFUs as discussed in Section \ref{sec:performance}.  The effective resolution with the IFU is 25\% higher than with a 1$\asec$ slit as described in Section \ref{sec:science}.  The Binospec IFU is particularly useful for measuring the kinematics and stellar population of galaxies compatible with its 12$\asec$ by 16$\asec$ format. 
A fully functioning pipeline is available to produce complete processed datacubes.

\begin{acknowledgments}
We thank FiberTech Optica for their work mounting the optical fibers during the challenges of the pandemic.  We thank Will Goble and the MMT staff for their help commissioning the Binospec IFU.  We thank Stephen Shectman, the referee, for a careful reading and helpful comments.
\end{acknowledgments}

\vspace{5mm}
\facilities{MMT(Binospec)}


\bibliography{BinoIFU}{}
\bibliographystyle{aasjournal}

\end{document}